\documentclass[prl,aps,amssymb,twocolumn,superscriptaddress,notitlepage]{revtex4-1}
\usepackage{amsmath}
\usepackage{amssymb}
\usepackage{amsthm}
\usepackage{amsfonts}
\usepackage{listings}
\usepackage{enumerate}
\usepackage{latexsym}
\usepackage{psfrag}
\usepackage{calrsfs}
\usepackage{bm}
\usepackage{graphicx}
\usepackage[caption=false]{subfig}
\usepackage{array}
\usepackage{color}
\usepackage[normalem]{ulem}
\usepackage{hyperref}

\def\bk{{\sf k}}
\def\bA{{ A}}
\def\bB{{\sf B}}

\def\bM{{ M}}

\def\bR{{R}}

\def\up{{\uparrow}}
\def\dn{{\downarrow}}

\def\T{\mathcal{T}}

\def\C{\mathcal{C}}
\def\P{{\sf P}}

\def\H{\mathcal{H}}

\def\K{\mathcal{K}}

\def\inv{^{-1}}
\def\half{{1\over2}}

\def\bR{{\sf R}}

\newcommand{\beq}{\begin{equation}}
\newcommand{\eeq}{\end{equation}}
\newcommand{\beqarray}{\begin{eqnarray}}
\newcommand{\eeqarray}{\end{eqnarray}}

\newcommand{\bsigma}{\mbox{\boldmath$\sigma$}}
\newcommand{\ket}[1]{|#1\rangle}
\newcommand{\ev}[1]{\langle #1\rangle}

\allowdisplaybreaks

\begin{document}
\title{ Splitting the hinge mode of higher-order topological insulators}
\author{Raquel Queiroz}
\email{raquel.queiroz@weizmann.ac.il}
\affiliation{Department of Condensed Matter Physics,
Weizmann Institute of Science,
Rehovot 7610001, Israel}

\author{Ady Stern}
\email{adiel.stern@weizmann.ac.il}
\affiliation{Department of Condensed Matter Physics,
Weizmann Institute of Science,
Rehovot 7610001, Israel}

\date{\today}
\begin{abstract}
The surface of a higher order topological insulator (HOTI) comprises a two-dimensional topological insulator (TI) with broken inversion symmetry, whose mass is determined by the microscopic details of the surface such as surface potentials and termination. It hosts a helical mode pinned to selected hinges where the surface gap changes its sign. We study the effect of perturbations that break time-reversal and particle-conservation on this helical mode, such as a Zeeman field and a proximate superconductor. {We find that in contrast to the helical modes of inversion symmetric TIs, which are gapped by these couplings, the helical modes at the hinges can remain gapless and spatially split.} When this happens, the Zeeman field splits the helical mode into a chiral mode surrounding the magnetized region; and a superconductor results in a helical Majorana mode surrounding the superconducting region. The combination of the two might lead to the gapping of one of the chiral Majorana modes, and leave a single one-dimensional chiral Majorana around the superconducting island. We propose that the different topological states can be measured in electrical transport.
\end{abstract}
\maketitle

\paragraph{Introduction.---}
Three dimensional time-reversal invariant higher-order topological insulators (HOTIs) have been predicted to host  protected helical modes in their one dimensional hinges \cite{Benalcazar17a,Benalcazar17b,Schindler18a,Schindler18b,Song17,Khalaf17,Khalaf18,Langbehn17,Geier18,Imhof2017,Peterson18,SerraGarcia18}.  Promising candidates are Bismuth \cite{Schindler18b} and strained
{\rm SnTe} \cite{Schindler18a,Hsieh12}.
Together with crystalline  \cite{Fu11,Hsieh12}, and weak topological insulators, HOTIs rely on crystalline symmetries of the bulk to protect their surface modes. In fact, HOTIs may be viewed as topological crystalline insulators in which the surface breaks the symmetry that protects the surface modes \cite{Khalaf17,Khalaf18}. This leads to a (possibly small) mass gap making the surface low-energy resemble a two-dimensional topological insulator (TI) without inversion symmetry, akin to a thin-film three-dimensional TI \cite{Yu61,zhang2010crossover,PhysRevB.81.115407,shan2010effective}.
A single helical mode can be localized at a hinge connecting faces with a surface gap of opposite sign. In intrinsic HOTIs, this domain wall is required to exist by the bulk symmetry through global constraints on the mass function \cite{Schindler18a,Khalaf17,Geier18}. However, at a local level, the protection of the helical mode relies only on time-reversal and charge conservation symmetries. Therefore, the hinge mode will not back-scatter in the presence of (for example) a sharp turn.

\begin{figure}[t!]
    \centering
    \includegraphics[width=\columnwidth]{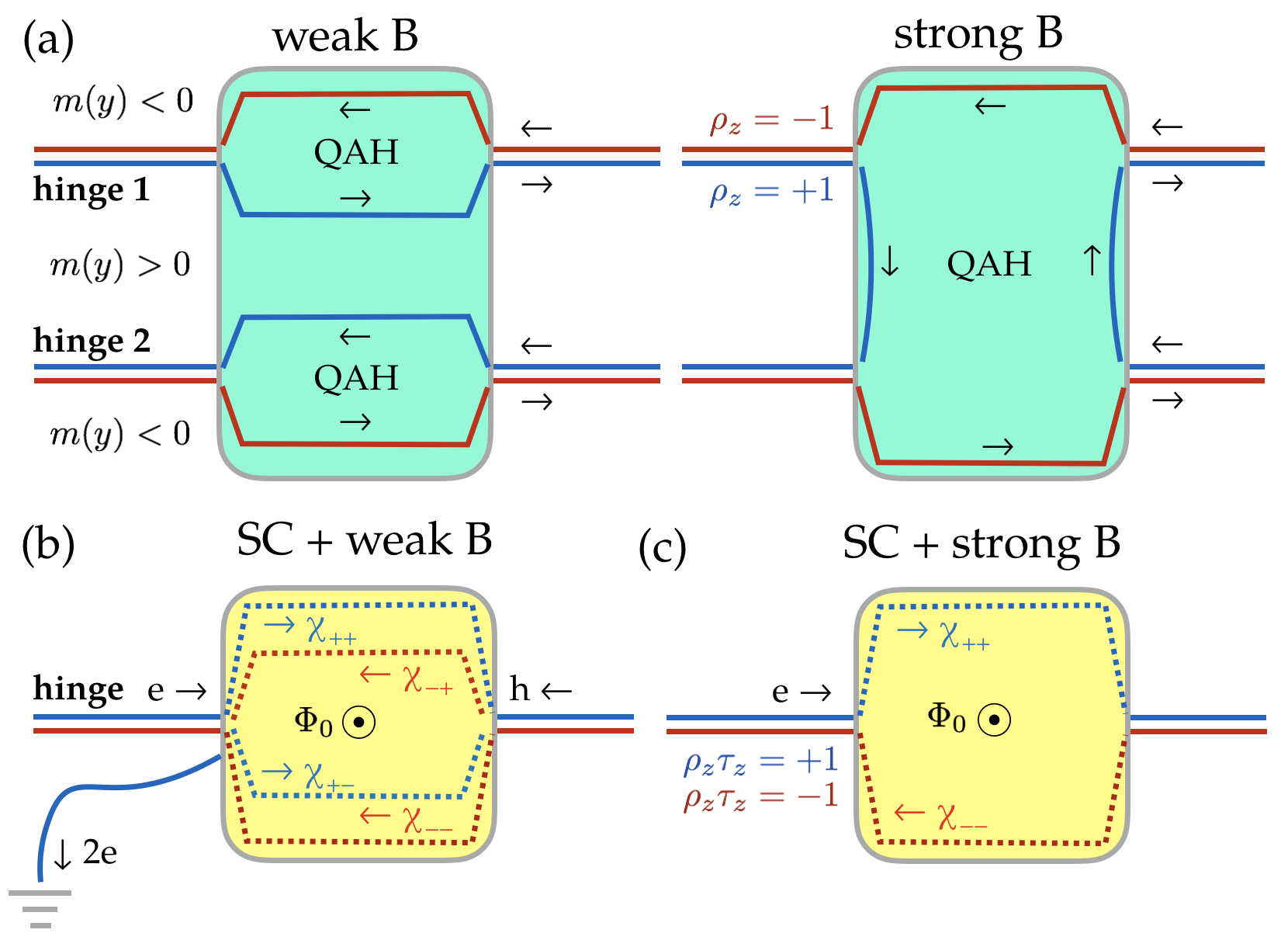}
    \caption{(a) A Zeeman field splits the hinge state into two spin-momentum locked chiral modes, forming a quantum anomalous Hall (QAH) region. For a strong enough field, two chiral modes from different hinges overlap extending the QAH over the entire region in between hinges. (b) Proposed Majorana interferometer, as an extension to Refs. \cite{Fu09b,Akhmerov09}. A magnetic flux $\Phi_0$ is enclosed by a helical Majorana mode. The interference will be evident by an electron current from the superconductor to the ground, that ensures charge conservation. (c) When one chiral mode is gapped, the flux is encircled by a single chiral Majorana mode. It realizes the junction proposed in Refs.\cite{Qi10,Chung2011,Wang2015,Huang17} where the conductance is quantized to $e^2/2h$.}
    \label{fig:experimental_setup}
\end{figure}

In this work, we study the fate of the hinge modes when subjected to perturbations that break time-reversal or charge conservation through exposure to an external magnetic field and proximity to a superconductor. 
We find that while helical modes on the edges of inversion symmetric two-dimensional TIs are gapped by these perturbations \cite{Fu08,Fu09a,Teo10a,Teo10b,Fu09b,Akhmerov09}, it is generically not true for helical modes in HOTIs. Instead, we find it is possible that the helical mode remains gapless but spatially split. 
The origin of this extended map of possibilities is two-fold: First, the surface gap reflects the local breaking of a bulk symmetry, which can be small and controlled by surface perturbations and (or) orientations. The hinges of the material realize a natural domain wall in the mass function which may vary smoothly over a length scale larger than the lattice constant. Second, the surface of a HOTI is strongly spin split by Rashba spin-orbit coupling, which implies a momentum mismatch between the low-energy Dirac valleys. Such mismatch makes the Zeeman field act mostly within each valley, crucial to guarantee the helical mode is split rather than gapped.

The helical fermionic mode can be split into two chiral fermions by a Zeeman field: The area confined by these modes becomes an effective surface-based 
Chern insulator with the spin of the two chiral modes oppositely polarized; Alternatively, by a combination of Zeeman and superconductivity it can be split into four chiral Majorana modes: The area confined between them forms a surface-based helical or chiral topological superconductor, the latter by letting two of the chiral Majoranas gap each other, see Fig.\ref{fig:experimental_setup}.
These different scenarios can be distinguished by charge transport. We extend the previously proposed Majorana interferometer on the surface a three dimensional TI \cite{Fu09b,Akhmerov09} to the surface of HOTIs.

\paragraph{ Model.---}The surface model we consider originates from the topology of a three dimensional bulk: If the crystalline symmetry is preserved at the surface, the surface gap vanishes and it realizes the anomalous Dirac theory of a TCI with two surface Dirac cones (valleys) \cite{Ando15,Liu13,Serbyn14,Nielsen81,Fang17}. Generally, these valleys will be separated in momentum. The two valleys are gapped by a time-reversal invariant mass, which can be interpreted as an applied magnetic field with an opposite sign at each valley \cite{Schindler18a,Khalaf17,Khalaf18}. Typically other surface and bulk states are at higher energies, thus the surface is well described by a continuum Dirac Hamiltonian. This is the case we consider. With $\H\!=\!\Psi^\dag H \Psi$ and $\Psi\!=\!((c_{\up+},c_{\up-}),(c_{\dn+},c_{\dn-}))^T$ the fermionic operators carrying a spin index $\sigma_z=\up,\dn$ and a valley index $\rho_z=+,-$, the Hamiltonian is given by
\begin{align}
H=
v\bk\cdot\bsigma\rho_0+\bB_v\cdot\bsigma\rho_z\label{eq:HHOTI1}.
\end{align}
Here,
$\bk\!=\!(k_x,k_y,0)$ and $\bsigma\!=\!(\sigma_x,\sigma_y,\sigma_z)$ represent the surface momenta and the vector of spin matrices. 
The two valleys have the same helicity and velocity $v$, but feel an opposite effective field $\bB_v\!=\!(vk_x^0,vk_y^0,m)$. Here $\bk^0$ is the momentum separation with magnitude $k_0$ between the valleys and $m$ the surface gap. $H$ is symmetric under time reversal $\T$ which interchanges the valleys, taking the form $\T\!=\!i\sigma_y\rho_x\K$, with $\K$ complex conjugation.  
As for Weyl semimetals \cite{Wan2011}, the momentum separation of the two valleys makes  $\rho_z$ an (approximate) constant of motion, preserved by perturbations that vary slowly on the scale of $1/k_0$. Inversion acts on Eq.\eqref{eq:HHOTI1} as ${\cal P}=\sigma_z\rho_z$, thus the low-energy Hamiltonian explicitly breaks it, except when $k_0\!=\!0$. In this case Eq.\eqref{eq:HHOTI1} describes the low-energy theory of an inversion symmetric TI close to a phase transition \cite{Hasan10,Qi11,Kane05,Bernevig06,Fu07} in a basis that mixes spin and valley subspaces \cite{SM}.  

\paragraph{Hinge modes.---} 
Eq.\eqref{eq:HHOTI1}
allows for a single helical state localized at one-dimensional domain walls where $m(y)$ changes sign. In a HOTI, this is expected to happen where the surface changes orientation, its hinges. 
Here, we analyze a single hinge located at $y=0$ where $m(0)\!=\!0$ and $m(+\infty)\!>\!0$. With an eye to adding new mass terms to the Dirac Hamiltonian, we consider a generalized ansatz for the hinge mode wavefunctions \cite{Jackiw76},
\begin{align}
\psi(y,k_x)=P\Omega(y)\chi(k_x)/N, \quad P\chi(k_x)=\chi(k_x),\label{ansatznosc}
\end{align}
which
satisfies the Schr\"odinger equation $H\psi(y,k_x)\!=\!E(k_x)\psi(y,k_x)$.
Here, $N$ is a normalization constant, $P$ a projector that selects the eigenvectors of $\Omega(y)$ that decay exponentially at $|y|\to\infty$, and $\chi(k_x)$ is an eigenstate of the projected hinge Hamiltonian $H_e\!=\!PHP$. Substituting $k_y\to -i\partial_y$ in Eq.\eqref{eq:HHOTI1}, we find that $\Omega(y)$ satisfies
\begin{align}\textstyle
    (-i\Gamma\partial_y+{1\over v}\bM(y)+e{\bA})\Omega(y)\!=\!0,\quad [\Omega(y),H_e]\!=\!0,  \label{eq:diffeqomega}
\end{align}
where we have collected the terms of the Hamiltonian projected out by $P$ into gapping terms, $M(y)$ which respect $\{M(y),\Gamma\}\!=\!0$, and additional terms $A$ satisfying $[A,\Gamma]\!=\!0$. The former include the surface gap $m(y)$, as well as the Zeeman and superconducting terms to be introduced below. 
Akin to an inplane magnetic field on the surface of a 3D TI \cite{Sitte12}, terms collected in $A$ can be absorbed by a gauge transformation, provided they are simultaneously diagonalizable with $M(y)$ and $\Gamma H_e$. This is the case for $eA\!=\!k^0_y\sigma_y\rho_z$, the valley momentum-separation perpendicular to the edge. 
In our convention, $P$ is fixed to select the positive eigenvalues of $i\Gamma\bM(\infty)$.  Here, $\Gamma\!=\!\sigma_y$ and $M(y)\!=\!m(y)\sigma_z\rho_z$ imply that $P\!=\!(1+\sigma_x\rho_z)/2$ and $H_e(k_x)\!=\! v(k_x\!+\!k_x^0\rho_z)\sigma_xP$, which includes only those terms that commute with the projector. The effective Hamiltonian $H_e(k_x)$ admits two nonzero eigenstates $\chi_s$ labelled by their chirality $s\!=\!\pm1$, with energies $E_s(k_x)\!=\!svk_x+vk_x^0$. They take the explicit form $\chi_-\!=\!((0, -1),(0, 1))^T$, and $\chi_+\!=\!((1, 0),(1, 0))^T$. The edge orientation determines the spin quantization axis, and the projector correlates the eigenvalues of the valley and spin degrees of freedom of each mode.
From Eq.\eqref{eq:diffeqomega} $ \Omega(y)$ is easily obtained,
\begin{align}
\textstyle\Omega(y)\!=\!\exp\{-i\Gamma(e\bA y+{1\over v}\int_{0}^{y} \bM(y') dy')\}.\label{eq:solomega}
\end{align}
If we consider
$m(y)\!=\!m\tanh(y/y_0)$, with $m$ a positive constant, and $y_0$ determining the sharpness of the domain wall. Then the hinge modes acquire the simple form $\psi_s(y)\!=\!\Omega_s(y)\chi_s$ with
\begin{align}
    \Omega_s(y)=\exp\{sik_y^0 y\}({\rm sech}~ y/y_0)^{m y_0\over v}.\label{eq:solomegahinge}
\end{align}
In the neighboring hinge, the sign of $m$ is reversed, and the projector $\bar P\!=\!(1-\sigma_x\rho_z)/2$ ensures that the eigenvalues of $\rho_z$ and $\sigma_x$ are opposite to one another. The modes at this hinge are $\bar\psi_s(y)=\Omega_{-s}(y)(\rho_x\chi_s)$.

\paragraph{Zeeman field.---}
We now consider applying an external Zeeman field in the $z$-direction, either by a magnetic field or by proximity to a ferromagnet. 
{In the surface of a HOTI 
the $2k_0$ distance between the Dirac cones implies that a slowly varying Zeeman field will act diagonally in the valley subspace, $H+\bB\cdot\bsigma$.}
The only Zeeman term that can gap the hinge mode is $B^z\sigma_z$, perpendicular to the surface. It \emph{commutes} with the surface mass $m(y)$, hence changing the magnitude of the effective mass of each valley. 

To find how the Zeeman field affects the hinge mode, we apply the ansatz in Eqs.\eqref{ansatznosc} to \eqref{eq:solomega} with an effective mass 
$\bM(y)\!=\!m(y)\sigma_z \rho_z+B^z(y)\sigma_z$. Provided that far away from the hinge $m(y)$ remains non-zero but the magnetic gap vanishes. It is $m(y)$ that
defines the projector, and $P$ remains unchanged. Rewriting $i\Gamma\bM(y)\!=\![\sigma_x \rho_zm(y)+\sigma_xB^z(y)]$, we see that one valley feels a local increase in the surface gap, while the other a decrease. Consequently, the point where the two chiral modes are located is no longer the same, and become separated in $y$: they are localized at the two $y$ values for which  $m(y)\!=\!\pm B^z(y)$. Assuming symmetry around $y\!=\!0$, we denote these values $y\!=\!\pm y_Z $, where the sign is determined by $\sigma_x$. With the effective Hamiltonian unchanged, the chirality is also determined by $\sigma_x$. Hence, 
the chirality is locked to the $y$ position, as expected for a Chern insulator, see Fig.\ref{fig:experimental_setup}(a). Interestingly, the spins of the two counter-propagating chiral modes are polarized in opposite directions. 
The modified shape of the wavefunctions can be calculated if the Zeeman mass function is expressed by 
$B^z(y)\!=\!B^z{\rm sech~}2y/y'_0 $, confined in a region around the hinge, 
\begin{align}\Omega^Z_{s}(y)=\Omega_{s}(y) \exp\{sB^z{ y'_0\over v} \arctan\tanh \frac{y}{y'_0 }\},\end{align}
with $\Omega_s(y)$ defined in Eq.\eqref{eq:solomegahinge}. This shift will be noticeable provided $y_0$ is large enough. The vectors $\chi_s$ are not changed by $B^z$. 

Now we comment on Zeeman terms that scatter between Dirac cones and may gap the counter-propagating modes, such as terms proportional to $\sigma_z\rho_x$ or $\sigma_z\rho_y$. These are expected to be weak since they involve scattering to states at an energy of the order of $vk_0$, which we assume larger than the Zeeman gap and the surface gap. { At the hinge these terms can gap the helical mode.} However, this effect is suppressed by the spatial separation between the two chiral modes. The inplane fields $B^x\sigma_x$ and $B^y\sigma_y$ result in a joint shift in the location of the two valleys, but does not gap the hinge modes. This is particularly relevant when an external magnetic field induces the Zeeman field on the hinge. When two surfaces meet at an angle, an in-plane field is unavoidable. 

Let us consider a Zeeman mass that extends over a region with two hinge modes, localized at $y\!=\!\pm y_h$. Here, two different situations may occur. For a weak field, the helical modes of the two hinges will each split, and the chiral states will be localized at $y\!=\!\pm y_h\!\pm\! y_Z $. 
As the Zeeman mass gets stronger and $y_Z $ approaches $y_h$, two of the modes will get close to one another. With different values of $\rho_z\sigma_x, $ associated with the two hinges and the rigid locking of the spin $\sigma_x$ to the velocity in the $x$-direction, the spins of these counter-propagating modes will be anti-parallel and have the same velocity. 
Once the two modes have spatial overlap, the Zeeman field directed at the $z$-direction will couple the oppositely polarized spins and gap the two modes. The entire region between the hinges becomes a quantum Hall anomalous state, bounded by two chiral modes, each originating from a different hinge. 

\paragraph{Superconductivity.---}We now consider an additional coupling of the region around one hinge to a superconductor, such that by proximity it induces pairing at the surface. 
We find that both a helical and a chiral topological superconducting phases are possible, the latter under proximity with a single-band $s$-wave superconductor.

Superconductivity is introduced at the mean field level by adding a particle-hole subspace $\tau_z$. Following the convention of Ref.~\cite{Fu08}, we write the Bogoliubov-de Gennes (BdG) Hamiltonian $\H\!=\!\Phi^\dag \bar H\Phi/2$ with $\Phi^\dag\!=\!(\Psi^\dag,\Psi^TU_T)$ and $U_T\!=\!i\sigma_y\rho_x$. This Hamiltonian has built-in particle-hole symmetry $\C\!=\!i\tau_yU_T\K$ that anti-commutes with $\bar H$. In the Supplementary Information (SI) \cite{SM} we explicitly derive from coupling the surface Hamiltonian to a superconductor the possible induced pairing terms and their relative magnitudes. There are two singlet superconducting terms which emerge with comparable magnitudes: $\Delta_0$ which pairs across valleys, and $\Delta$ pairing within a valley. For a single-band $s$-wave superconductor, we find $\Delta_0\!\ge\!\Delta$, with the limiting condition $\Delta\!=\!\Delta_0$ satisfied when Cooper-pairs tunnel into both valleys indiscernibly \cite{SM}. 

We now write the BdG Hamiltonian with both superconductivity and the valley-preserving Zeeman field,
\begin{align}
    \bar H=v\bk\cdot\boldsymbol{\sigma}\tau_z+\bB_v\cdot\boldsymbol{\sigma}\rho_z\tau_z+\bB\cdot\boldsymbol{\sigma}+
    (\Delta_0+\Delta\rho_x)\tau_x\label{eq:HSCHOTI}.
\end{align}
Pairing is taken to affect a single hinge, assuming that the neighboring ones are far away \cite{Loss}. We now ask under which circumstances the helical mode  is split to form either helical or chiral Majorana modes, rather than being gapped. 
A necessary condition for a single gapless one-dimensional Majorana mode (helical or chiral) to exist is that there is a set of real and spatially {uniform} parameters $\bB_v$ where the BdG Hamiltonian admits a zero eigenvalue at $\bk\!=\!0$. Clearly, this is the case when the mass, Zeeman field and superconductivity all vanish. We look, however, for other combinations. Noting that at $\bk=0$, $\Delta\rho_x\tau_x$ commutes with all terms in the Hamiltonian, we are left to diagonalize the remaining terms.
We find that the needed combination is 

\begin{align}
\Delta^2=B^2+B_v^2+\Delta_0^2\pm2\sqrt{(\bB\cdot\bB_v)^2+(\Delta_0B)^2},\label{cond}
\end{align}
and $\Delta$, $\Delta_0$ and $\bB$ are controlled externally. Let us focus on some simplifying cases. 
With time-reversal symmetry $\bB\!=\!0$, it follows that only when $\Delta^2>\Delta_0^2$ there might be real solutions for $\bB_v$. If $k_0$ does not change in the region around the hinge, a real-valued solution for $m$ requires $\Delta^2>\Delta_0^2+v^2k_0^2$. When this condition holds, the helical fermion is split into two helical Majorana modes, rather than being  gapped. This does not happen when the superconductor is a single band $s$-wave superconductor, see SI. Splitting the helical mode in a time-reversal symmetric fashion requires the contacted superconductor to have additional structure in its order parameter, in agreement with Ref.\cite{Haim16}. 

With a nonzero Zeeman field, the situation is  different. If the edge modes are not gapped, we can have either two or four solutions at different values of the mass, corresponding to a possible split of the helical fermionic mode into either two or four chiral Majorana states at different positions. Eq.\eqref{cond} can be satisfied in different regimes. When the fields are collinear $\bB\cdot \bB_v=BB_v$, Eq.\ref{cond} implies that $\Delta^2>(\Delta_0\pm B)^2$. The threshold for $\Delta$ is reduced linearly with $B$ and becomes more accessible. If it happens that $k_0$ is significantly reduced around the hinge, for example when global constraints impose a domain wall in this term, then a Zeeman field of smaller magnitude than the induced gap can push the system through a topological phase transition into a chiral $p+ip$ superconductor, with split hinge modes. If, on the other hand, we consider the field perpendicular to the surface, $\bB\cdot \bB_v\!=\!B^zm$. Here it follows that a real-valued mass function $m$ implies that $\Delta^2\!>\!(\Delta_0\pm B^z)^2+v^2k_0^2$. 

Analogously to what we did in the previous section, we derive the spatial profile of the wavefunctions with an out-of-plane field, $B^z$, but also under the proximity of a superconductor that induces both $\Delta$ and $\Delta_0$ in comparable magnitudes. To apply the ansatz in Eq.\eqref{ansatznosc}, we require a hierarchy of energy scales: At $y\to\pm\infty$ the largest energy scale is determined by $\bB_v$, and thus $\bB_v$ determines the projector $P$, and it remains unchanged from the unperturbed hinge mode. Both a magnetic field $B^z(y)$ and superconductivity with $\Delta(y)$ and $\Delta_0(y)$ are added in a region around the hinge. At the hinge there are two oppositely propagating chiral fermionic modes (four chiral Majorana modes) located at $y_h$. The projected BdG Hamiltonian preserves only terms which commute with $P$. All other terms are exponentially suppressed by the distance between hinges. The effective BdG Hamiltonian reduces to $\bar H_e=[vk_x\sigma_x\tau_z+vk^0_x\sigma_x\rho_z\tau_z+\Delta_0\tau_x]P$, which has a gapped spectrum $E(k_x)=\pm\sqrt{v^2(k_x\pm k^0_x)^2+\Delta_0^2}$.  However, there is another possibility: the four chiral Majorana modes can be spatially separated from one another, in which case the matrix elements induced by $k^0_x$ and $\Delta_0$ become exponentially suppressed.  Interestingly, here $\Delta$ plays the role $B^z$ played previously. Under the conditions discussed above we can solve for the bound states of Eq.\eqref{eq:HSCHOTI}, assuming $k^0_x=\Delta_0=0$, and then include them perturbatively. All other terms $m\sigma_z\rho_z\tau_z$, $B^z\sigma_z$ and $\Delta\rho_x\tau_x$ commute with each other, and the wavefunctions can be directly calculated with Eq.\eqref{ansatznosc} taking 
$i\Gamma M(y)\!=\![\sigma_x \rho_zm(y)+\sigma_x\tau_zB^z(y)+\sigma_y\rho_x\tau_y\Delta(y)]$. With this, we find it convenient to label the four eigenstates of the hinge BdG Hamiltonian $\chi_{sl}$, where $s$ is the eigenvalue of the chirality operator $\sigma_x\tau_z\chi_{sl}\!=\!s\chi_{sl}$, and $l\!=\!\pm1$ is an additional quantum number $\sigma_y\rho_x\tau_y\chi_{sl}\!=\!l\chi_{sl}$, which distinguishes the Majorana modes. The effective mass term has four distinct zeros $y_{sl}\!=\!y_h\!+\!sy_Z\!+\!ly_\Delta$, which correspond to spatial location of the four chiral Majorana modes. The states $\chi_{sl}$ are explicitly given in terms of $\chi_s$ by $\chi_{sl}\!=\!(\chi_{s},-sl\chi_{-s})^T$. An explicit form of $\Omega(y)$ can be found in the limit where $\Delta(y)\!=\!\Delta{\rm sech~}2y/y'_0$, and is given by
\begin{align}\Omega_{sl}(y)\!=\!\Omega_{s}(y)\exp\{(sB^z\!+\!l\Delta){ y'_0\over v} \arctan\tanh \frac{y}{y'_0}\}.\end{align}
We now look at the matrix elements of the momentum shift $k_x^0$ and gap $\Delta_0$ in the Hamiltonian $\bar H_e$.  First, 
$k^0_x$ couples solutions of the same chirality but opposite  $l$, $\ev{\chi_{sl}|\sigma_x\rho_z\tau_z|\chi_{s'l'}}\!\propto\!\delta_{ss'}\delta_{l,-l'}$. 
Here $\delta$ is the Kronecker delta. This term will shift the energy of the Majorana modes but not gap them. 
On the other hand, $\Delta_0$ introduces a gap in the spectrum within one hinge, since the matrix elements $\ev{\chi_{s,l}|\tau_x|\chi_{s'l'}}\!\propto\!\delta_{s',-s}\delta_{l',-l}$ couple opposite chiralities and $l$, independently of its strength. 

The Zeeman field $B^z$ and paring $\Delta$ act as tuning knobs that determine the location of $\chi_{sl}$. There are four distinct and interesting scenarios. First, all $\chi_{sl}$ are at the same $y$. The edge spectrum is fully gapped, with the fermionic modes separated in momentum by $2k_0$. 
Second, all four $\chi_{sl}$ are separated from one another. The Hamiltonian reduces to $\bar H_e\chi_{sl}=svk_x\chi_{sl}$ reflecting four topologically robust chiral Majorana modes. We can also consider bringing two modes close  together and let them interact. If these states have the same chirality but different $l$, then they will remain gapless but shifted in momentum. If the states have opposite $s$ and $l$, they are gapped by $\Delta_0$: Two spatially separated chiral Majorana modes are left, surrounding a surface-based $p\!+\!ip$ topological superconductor. Lastly, in the case of time-reversal symmetry, $B^z\!=\!0$, the two helical Majorana modes of fixed $l$ are separated from each other surrounding a surface-based helical topological superconductor.

\paragraph{Transport.---} The split Majorana helical modes can be probed by a generalization of the chiral-Majorana-fermion interferometer proposed by Fu and Kane \cite{Fu09b}, and by Akhmerov, Nilsson and Beenakker \cite{Akhmerov09}. In that interferometer the surface of a strong TI is gapped by ferromagnets of opposite magnetization, separated by a superconducting island. In our set-up, shown in Fig.\ref{fig:experimental_setup} (b) we have two copies of this interferrometer, of two different chiralities. When a voltage is applied between two points on the hinge mode on two sides of a grounded superconductor, current will flow in one of these chiralities. The Majorana modes form a closed loop surrounding the superconducting region, where we can pierce a magnetic flux quanta $\Phi_0\!=\!h/2e$. In analogy to Refs.\cite{Fu09b,Akhmerov09}, the Majorana modes acquire a phase, which will turn an incident electron into a hole. In this case, the electron current is converted into a hole current, and, by charge conservation, creates an electric current of twice the incident current into the superconductor; When, alternatively, two of the four chiral modes are gapped, the magnetic flux is enclosed by a single chiral Majorana mode, and a vortex traps a single localized Majorana zero mode. This device is analogous to the QAH-$(p+ip)$-QAH juntion proposed by Zhang \emph{et.al.} in Refs. \cite{Qi10,Chung2011,Wang2015,Huang17}. In the absence of back-scattering, this device has a quantized conductance of $e^2/2h$, reflecting the fact that a chiral Majorana mode for each incident electron is fully reflected.

\paragraph{Conclusion.---} We studied the conditions under which hinge modes of higher-order TIs 
shift away from the hinge, in contrast with the expected gapping. When coupled to a Zeeman field the hinge mode can split into two one-dimensional modes of opposite chirality, while an additional superconductor may further split it into four chiral Majorana one-dimensional modes.
Our observations open the way for a generalization of the Majorana interferometer, in which the basic properties of the neutral Majorana modes find their way to charge transport. Furthermore, our results highlight the possible manipulation of hinge modes based on their coupling to ferromagnets and superconductors.  

\paragraph{Acknowledgments}
{ The authors thank Roni Ilan, Andrei Bernevig and the participants of the CRC183 miniworkshop on topological phases with
higher-order boundary states, in particular Eslam Khalaf, for insightful discussions.  This work was supported by the Israel Science Foundation; the European Research Council under the Project MUNATOP; the DFG (CRC/Transregio 183, EI 519/7-1).}

\bibliography{MyCollection}
\clearpage

\onecolumngrid

\section{Supplementary Information for\\"Splitting the hinge mode of higher-order topological insulators"}

This supplementary information is organized as follows: In Sec.A, we complete the discussion of the main text with a tight-binding Hamiltonian where our results can be simulated numerically. The model we present corresponds to a two-dimensional lattice model of an inversion-breaking topological insulator. 
We study the phase diagram of the lattice model under two uniform perturbations: A Zeeman field and superconductivity, by varying the effective valley field $\bB_v$. We explicitly add terms that gap the hinge spectrum and show that due to the splitting, the gap introduced by these terms becomes exponentially suppressed.  In Sec.B we study the splitting of the edge mode at a domain wall by artificially including an inhomogeneous mass function that varies smoothly over a few lattice sites. We contrast it with the gapping of the same edge state when the mass function has a sharp discontinuity, as it can be seen on the opposite side of the sample. Around the smooth edge region, we add an island of a ferromagnet and a superconductor to observe the split into either chiral fermion or helical Majorana modes. In Sec.C, we study the origin of the different superconducting pairing terms considered in the main text under the proximity to an $s$-wave superconductor. We find that for a single-band superconductor in the case where pairing does not distinguish between the two valleys, the intra-valley and inter-valley pairing magnitudes satisfy $\Delta=\Delta_0$. We show that the condition $\Delta>\Delta_0$ can only be satisfied by proximity to a superconductor with additional structure to its order parameter, we show an example for a two-band $s$-wave superconductor that generates only $\Delta$. In Sec.D we comment on the relevance of our result to known HOTI materials, Bismuth and strained SnTe.

\subsection{A. Lattice model of an homogeneous inversion breaking topological insulator}

In the main text, we studied the effect of a proximate Zeeman field and superconducting pairing on the surface of a three-dimensional higher-order topological insulator for a continuum Dirac Hamiltonian. Now, we construct a lattice model with compatible low-energy physics by performing the substitution $vk\to \varepsilon\sin k$, where $\varepsilon$ has units of energy, and $ m\to (m-2t+t\cos k_x +t\cos k_y)$ for $-t$ the hopping energy. We consider in all numerical calculations $\varepsilon=t=1$.
The lattice model is given by
\begin{align}
\H=&\sum_\bk\Psi_\bk^\dag H(\bk)\Psi_\bk,\nonumber\\ H(\bk)=&~\varepsilon\left (\sin k_x\sigma_x+\sin k_y\sigma_y\right)+v\left (k^x_0\sigma_x\rho_z+k^y_0\sigma_y\rho_z\right)+(m-2t+t\cos k_x+t\cos k_y)\sigma_z\rho_z,
\label{latticemodel}\end{align}
acting on the basis vector  $\Psi_\bk=((c_{\bk\up+},c_{\bk\up-}),(c_{\bk\dn+},c_{\bk\dn-}))^T$.
Time-reversal symmetry acts as $\T=i\sigma_y\rho_x\K$ and has an effective valley symmetry $\rho_z$, which here can be interpreted as an orbital degree of freedom. The terms $vk_0^x$ and $vk_0^y$ correspond to staggered spin-orbit coupling terms that act locally within an orbital. At each $\rho_z$ sector, this model can be seen as the tight-binding model for a Chern insulator. It is important to note that up to first neighbor coupling, Eq.\eqref{latticemodel} is very general. Additional terms which do not commute with $\rho_z$ can be added without qualitatively changing our discussion: Upon a basis transformation, the valley can be again diagonalized such that the model above is recovered with different parameters. We choose this basis since it renders our results very intuitive.
However, the reader might find useful to apply the unitary transformation
\begin{align}\tilde U=\half\exp\{  i \pi \sigma_z\rho_y/4 \}(\rho_+ + \sigma_z\rho_-),\end{align} to transform the above model into a basis where time-reversal symmetry acts as $\T=i\sigma_y\K$. In this case, the valley is given by the eigenstates of $\sigma_z\rho_x$. Note that in this basis, the linearized Hamiltonian, expanded close to $\bk\approx 0$, becomes \begin{align}\tilde UH(\bk\approx0)\tilde U^\dag\!=\!v\bk\cdot\bsigma\rho_z+\!v\bk^0\times\bsigma\rho_y+m\rho_x .\end{align}
Then, inversion $\P=\rho_x$ is recovered when $k_0=0$, which enforces the two valleys to be degenerate and centered at a time-reversal invariant point.

Let us now gain additional insight into this model by transforming it into real space. On a square lattice, the Hamiltonian \eqref{latticemodel} is translated into
\begin{align}
    \H=&\sum_{ij}\Psi^\dag_iH_{ij}\Psi_{j},\nonumber\\
    H_{ij}=&[(m-2t)\sigma_z\rho_z+ v\left (k^x_0\sigma_x\rho_z+k^y_0\sigma_y\rho_z\right)]\delta_{\bR_i,\bR_j}+\half\sum_{\bR_a}\left[t\sigma_z\rho_z+i\varepsilon(\bR_i-\bR_j)\cdot\boldsymbol{\sigma}\right]\delta_{\bR_i-\bR_j,\bR_a},\label{eq:unperturbedH}
\end{align}
where the fermionic creation operators $\Psi_i=((c_{i\up+},c_{i\up-}),(c_{i\dn+},c_{i\dn-}))^T$ act on the lattice site $i$ located at $\bR_i$, and $\bR_a$ correspond to the nearest neighbour vectors $\bR_1=\hat x$, $\bR_2=-\hat x$, $\bR_3=\hat y$ and $\bR_4=-\hat y$. The valley $\rho_z$ reflects a local atomic orbital, which is \emph{not} invariant under time-reversal. For example, it can represent $\ket{\pm}=\ket{p_x\pm ip_y}$ states, or any pair of orbitals with nonvanishing angular momentum that get interchanged under $\T$. The momentum separation of the two orbitals, $v\bk^0$, corresponds to an on-site potential, a spin-orbit coupling term with origin in crystal fields. These terms are forbidden by inversion symmetry, but for a non-centrosymmetric system they are generally present. Focusing on a single orbital/valley, these terms behave as an applied in-plane Zeeman field. Together with the mass, we can say that each orbital is subjected to the effective field ${\sf B}_v=(vk_0^x,vk_0^y,m-2t)$.

\subsubsection{Zeeman field}
\begin{figure}[t]
    \centering
    \includegraphics[scale=0.37]{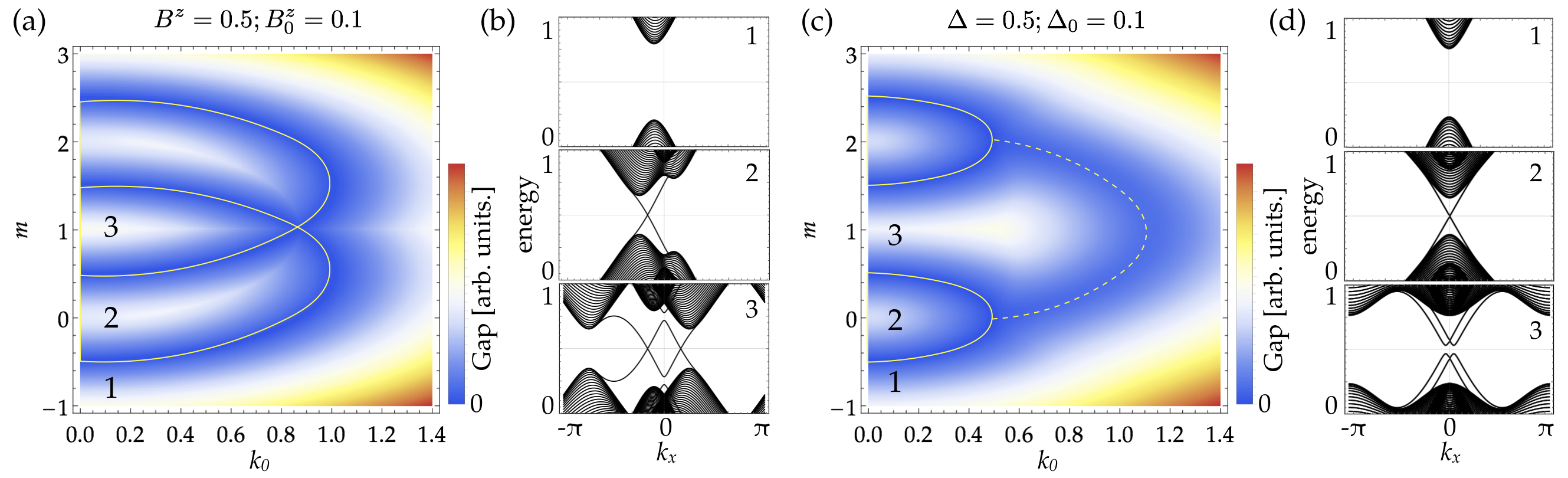}
    \caption{Phase diagram for homogeneous systems with an applied constant magnetic field ($B^z=0.5$, $B^z_0=0.1$)  (a,b); And superconductivity ($\Delta=0.5$, $\Delta_0=0.1$) (c,d). Panels (b) and (d) show a representative band structure of each phase, calculated with a ribbon geometry of 60 sites along $y$ with $k_0^y=0$, $m=(-1,0,1)$, and $k_0^x=k_0=0.5$ for (b) and $k_0^x=k_0=0.2$ for (d).
    (a,b) Zeeman field:  The phase diagram has three distinct regions, delimited by a solid yellow line that indicates a topological phase transition. Only the region 2 is topological:  a quantum anomalous  Hall state with chiral fermion modes at its edges. The phase labelled by 3 is protected by the separation of the two valleys in momentum space. At higher energies $\sim vk_0$ the two chiral branches  cross and will be gapped by $B^z_0$. This phase is protected by the approximate $\rho_z$ symmetry. 
 (c,d) Superconductivity: Up to a certain $k_0$ threshold, the phase diagram admits three distinct phases for varying $m$: trivial; time-reversal invariant topological superconductor with helical Majorana modes; and a topological superconductor protected by $\rho_z\tau_z$ which is broken by the pairing $\Delta_0$, resulting in gapped helical Majorana modes shifted in momentum. Their possible split into spatially separated helical Majoranas, robust to time-reversal invariant perturbations, can be confirmed by the existence of the topological regions adjacent to this phase. When $k_0$ is large, this is no longer possible. The dashed line indicates a gapped transition. The topological region (2) shrinks with increasing $\Delta_0$, disappearing exactly when $\Delta=\Delta_0$. }
    \label{fig:phasediag}
\end{figure}

In the main text, we have considered the effect of a Zeeman field that acts diagonally in valley space, $B^z$. We have shown that such term changes the position of the hinge mode, but does not lead to its gapping. In the numerical calculations we now show, we consider in addition the effect of a Zeeman field that couples the valleys, $B^z_{0}$. With inversion-breaking spin-orbit coupling, the effect of this term in the two-dimensional bulk can be neglected since the two bands are separated by an energy gap, and cannot be coupled through elastic scattering. However, the helical modes at the hinge/edge may be gapped by such term at crossings of where the two orbitals meet, which happens at energies $\sim vk_0$. We study the competition between the two in the lattice model, by adding to Eq.\eqref{eq:unperturbedH} the local terms
\begin{align}
    H_Z=B^z\sigma_z+B^z_0\rho_x\sigma_z.
\end{align}
The resulting Hamiltonian with uniform parameters will allow for a Chern insulating phase, provided the gap of one $\rho_z$ sectors is inverted. Fixing $B^z=0.5$, $B^z_0=0.1$ and $k_y^0=0$ we show in Fig.\ref{fig:phasediag} a phase diagram in function of $m$ and $k^x_0$. We show the magnitude of the gap at half filling. It is clear that there is a topological phase delimited by a parabolic region: In the absence of a Zeeman field, this region corresponds to a 2DTI phase with helical modes. Adding a nonvanishing $B^z$ shifts this parabola in an opposite fashion for each valley, thus creating two anomalous quantum Hall way with opposite Chern number adjacent to the 2DTI phase. Note that the helical modes in this 2DTI phase (region 3) are not protected, and are coupled by the $B^z_0$ term. The Chern insulating phases (region 2) has chiral modes. It is a topological phase when $\rho_z$ is an exact symmetry of the system. We present in panel (b) the band structure of representative systems in the three different phases for a ribbon geometry. 

\subsubsection{Superconductivity}

We now repeat the analysis above, this time including the two singlet superconducting terms considered in the main text, but with a vanishing Zeeman field. $\Delta$, which couples within an orbital; and $\Delta_0$ which couples the two orbitals. To the real space Hamiltonian Eq.\eqref{eq:unperturbedH}, we add the mean field pairing terms,
\begin{align}
\H_\Delta=\Delta_0(c^\dag_{i\up+}c^\dag_{i\dn-}+c^\dag_{i\up-}c^\dag_{i\dn+})+
\Delta(c^\dag_{i\up+}c^\dag_{i\dn+}+c^\dag_{i\up-}c^\dag_{i\dn-})+{\rm h.c.}.
\end{align}
In the Nambu basis $\Phi^\dag=(\Psi^\dag,\Psi^TU_T)$ these pairing terms assume the simple form presented in the main text, 
\begin{align}
    H_\Delta=\Delta\rho_x\tau_x+\Delta_0\rho_0\tau_x,
\end{align}
where the Pauli matrices $\tau_i$ act in particle-hole subspace. 

At this stage, we have not determined the origin of these coupling terms, and their amplitudes are simply set by hand. We have argued in the main text the hinge mode will split provided $\Delta>\Delta_0$. In the numerical calculations, we consider $\Delta_0=0.1$ and  $\Delta=0.5$. We show in Fig.\ref{fig:phasediag}(c) the phase diagram in function of both $m$ and $k_0^x$, while assuming $k_0^y=0$.
We find three distinct regions: A trivial region; a two-dimensional time-reversal symmetric topological insulator with helical Majorana modes; and a region with two helical fermions, gapped by $\Delta_0$ and thus not topologically robust. The dashed line indicates a gapped transition, where the gap vanishes when $\Delta_0=0$. In this limit $\rho_z\tau_z$ is an exact symmetry of the Hamiltonian which would protect a topological phase bounded by the dashed line. The topological regions (2) monotonically shrink with an increasing $\Delta_0$, disappearing for $\Delta=\Delta_0$.

\begin{figure}[t]
    \centering
    \includegraphics[scale=.37]{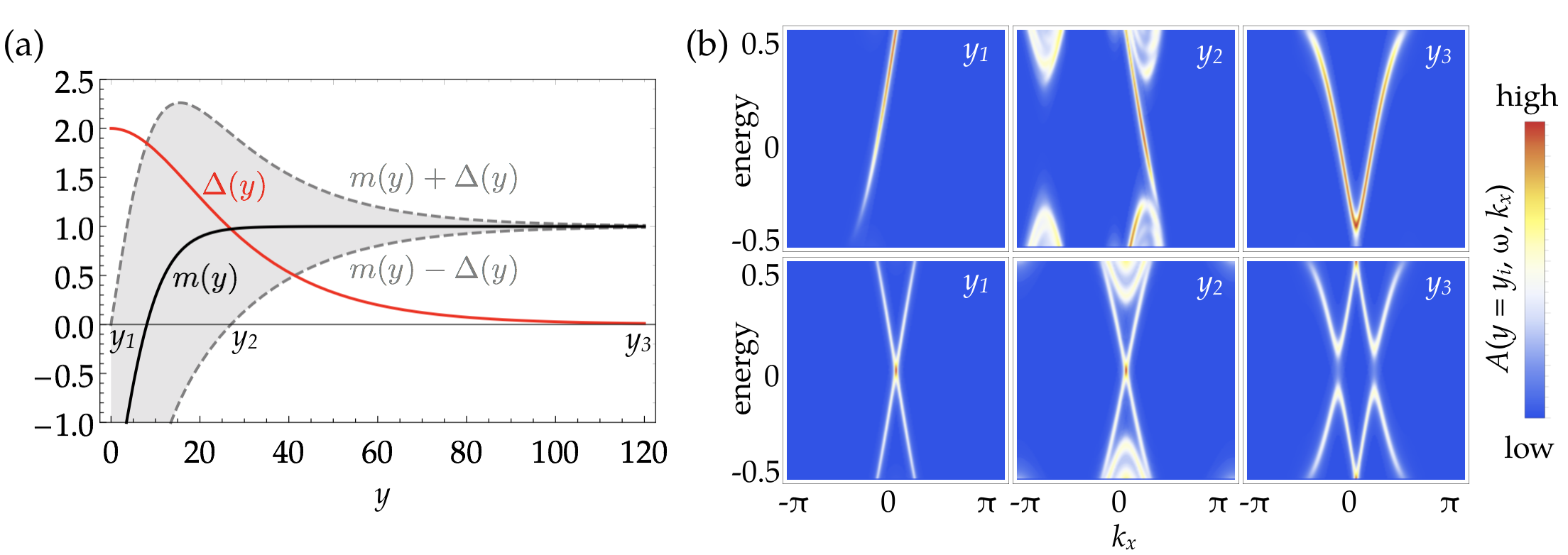}
    \caption{Numerical calculation for an inhomogeneous ribbon, where the $y$-dependent model parameters are shown in (a): On the left side the mass $m(y)$ changes smoothly, while on the right side the mass changes abruptly (back line). Either a Zeeman field $B^z(y)$ or superconductivity pairing $\Delta(y)$ is added in the region around the smooth edge (red line, only $\Delta(y)$ is shown to avoid cluttering).  The gray dashed lines show the sum and difference of the mass and the dominant perturbation, which have zeros at $y_1$ and $y_2$. These are the locations we expect the split modes to be. They delimit the QAH or the 2D topological superconducting regions. We additionally include a sub-leading perturbation, either $B^z_0$ or $\Delta_0$ with a constant amplitude and equal to $0.1$ across the entire ribbon. This is meant to compare its effect on the split modes on the smooth edge, with the spatially coexisting modes on the sharp edge. (b) We show the $y$-resolved spectral weight $A(y,\omega,k_x)$ for the ribbon geometry with non-homogeneous parameters. Zeeman field (top row): The helical fermion on the left edge is split into two chiral modes at $y_1$ and $y_2$.  The gapping introduced by $B^z_0$ is exponentially suppressed, and the region between chiral modes comprises a quantum anomalous Hall state. On the right edge, $y_3$, the helical mode is gapped by $B^z_0$ at the crossing located at $k_x=0$ and $\omega=0.5$. Superconductivity (bottom row): The helical fermion mode is split into two helical Majorana modes at $y_1$ and $y_2$, unperturbed by the gapping term $\Delta_0$. At the right edge, $\Delta_0$ gaps the Majorana modes split in momentum.} 
    \label{fig:nonhomogeneous}
\end{figure}

\subsection{B. Numerical simulation of a inhomogeneous mass profile}
The surface of a HOTI differs from a stand-alone two-dimensional topological insulator in the sense that a mass domain wall emerges naturally. The surface gap depends crucially on details such as termination, orientation and surface modifications to the electrochemical potential. There is a high degree of variation in these parameters, and the hinge represents the location where it changes sign. Importantly, the length scale on which changes in surface parameters happen are not necessarily bound to the lattice constant, in the same way they are in a bulk stand-alone system.
In contrast, in a two-dimensional bulk system, the bulk mass is an intrinsic parameter that we expect to abruptly change at the edge. 

In this section, we artificially include a slowly varying mass function over the two-dimensional tight-binding model and compare it directly with a sharp edge. This is shown in Fig.\ref{fig:nonhomogeneous}. In panel (a) we plot the mass profile $m(y)$ in black, smoothly varying on the left and sharply changing at the right edges, as well as the region where the main perturbation, either $B^z(y)$ or $\Delta(y)$ is added to the ribbon. The gray curves represent $m(y)\pm B^z(y)$ or $m(y)\pm \Delta(y)$, which are the effective mass at each one of the two valleys sees. In the entire sample, we include a small amount of either $B^z_0$ or $\Delta_0$, the perturbation that couple the two valleys. The region between comprises either an anomalous quantum Hall insulator or a 2D topological superconductor with helical modes. In panel (b) we show the $y-$ and $k_x-$resolved density of states at the zeros of the effective mass. We find the helical edge mode, which is gapped in the sharp edge, is either split into two chiral modes, localized at the lattice sites where $m(y)=\pm B^z(y)$ (top); or is split into two helical Majorana modes (bottom). 

In our numerical calculations, we have considered a ribbon of 120 sites along the $y$-direction and periodic boundaries along $x$.  
The spectral weight is calculated following Ref.\cite{Queiroz16},
\begin{equation} \label{Eq:intensity}
 A (y,\omega,k_x)=
-\frac{1}{\pi}\text{Im}\sum\limits_{m}\frac{\left| \psi_m(k_x,y)\right|^2}{\omega-E_m +i \eta},
\end{equation}
where  $\psi_m^{k_x,\nu}(y)$ is the $m$-th eigenstate of the ribbon Hamiltonian with energy $E_m$, and $\eta$ a small broadening taken to be $\eta=0.01$.

\subsection{C. Proximity effect to an $s$-wave superconductor}
In this section, we derive the magnitude of the superconducting terms by considering that the two dimensional, unperturbed, Hamiltonian Eq.\eqref{eq:unperturbedH} is put in contact with a superconductor described by the Hamiltonian
\begin{align}
    \H_{sc}=\sum_i[{\Delta}_{sc}]_{\mu\mu'} d^\dag_{i\mu}d^\dag_{i\mu'}+\rm h.c.,
\end{align}
with $\mu$ a combined spin and orbital index of the fermionic modes in the superconductor. 
For the sake of simplicity, we consider that the superconductor has only pairing energy but no kinetic energy. 
The Cooper pairs at the superconductor can tunnel to the surface or two-dimensional topological insulator with a hopping amplitude of
\begin{align}
    \H_{coup}=\sum_{\mu\rho}\lambda_{\mu\rho}d^\dag_{i\mu} c_{i\rho},
\end{align} 
where $\rho$ is the combined spin and orbital index of the fermionic modes in $\H$. We consider both the coupling $\lambda$ and the pairing $\Delta_{sc}$ to be strictly local, and thus their spatial indices can be omitted.
The combined Hamiltonian $\H_{full}=\H+\H_{sc}+\H_{coup}$ can be written as a Bogoliubov-de Gennes (BdG) Hamiltonian, $\H_{full}=\tilde\Phi^\dag\bar H_{full}\tilde\Phi$, by defining the Nambu basis vector $\tilde\Phi^\dag_i=(c_i^\dag,c_i^T U_T,d_i^\dag,d_i^TU_T^{ sc})$. Then, we can write
\begin{align}
    \bar H_{full}=\begin{pmatrix} \bar H&\bar \lambda\\\bar\lambda^T&\bar\Delta_{sc}{}
    \end{pmatrix},&&
    \bar H=\begin{pmatrix}H&0\\0&-H\end{pmatrix},&&\bar \Delta_{sc}=\begin{pmatrix}0&\Delta_{sc}\\\Delta_{sc}^\dag&0\end{pmatrix}, &&\bar \lambda=\begin{pmatrix}\lambda&0\\0&-\lambda \end{pmatrix}.
\end{align}
as in the main text, the bar indicates the BdG form. In this basis, the $s$-wave superconductor $\Delta_{\sc}$ is simply a constant multiplying the identity matrix in spin space.
We find the proximity-induced superconductivity by integrating out the superconductor's degrees of freedom. For this, we consider the Schur's complement
\begin{align}
\bar H_{eff}=\bar H-\bar\lambda^T(\bar\Delta\inv)\bar\lambda= H\tau_z+\Delta_{ind}\tau_x
\end{align}
acting in the reduced surface space $\Phi_i^\dag=(c_i^\dag,c_i^TU_T)$. Because of the simplified form we have chosen for the superconductor and coupling (only pairing), the induced term is purely a pairing term, proportional to $\tau_x$. Note that $\tau_y$ is forbidden by chiral symmetry. The induced pairing is explicitly given by \begin{align}\Delta_{ind}=-\lambda^T\Delta_{sc}\inv\lambda.\end{align} 

Now we look at the simplest case when $\Delta_{sc}$ is a single band $s$-wave superconductor, and $\Delta_{sc}\propto\sigma_0$ proportional to the identity matrix. A direct consequence of this is that $\Delta_{ind}\propto\lambda^T\lambda$ is necessarily a positive semi-definite matrix. Following Ref.\cite{Haim16}, a positive semi-definite $\Delta_{ind}$ does not allow for a topologically nontrivial phase. Therefore the condition imposed on the induced superconducting order parameter on the two-dimensional model will not overlap with the splitting condition derived in the main text.
Let us derive the induced pairing in this case. We assume that the Cooper pairs can tunnel into both valleys/orbitals with an amplitude $a$ and $b$, in a time-reversal invariant way. Then 
\begin{align}\lambda^T=\begin{pmatrix}b&0&a&0\\0&a&0&b\end{pmatrix},\end{align}
which results in
\begin{align}\Delta_{ind}={1\over\Delta_0} ({a^2+b^2\over 2}\rho_0+ab\rho_x+{a^2-b^2\over 2}\sigma_z\rho_z).\end{align}
It follows directly that the amplitude of the induced order parameter $\Delta_0\tau_x$ is always larger than the other two. This implies in particular that $\Delta>\Delta_0$ is not possible. When $a=b$ the two terms generated of equal amplitude correspond to $\Delta$ and $\Delta_0$. The third term implies spin and valley dependent coupling, and corresponds to a spin-triplet, time-reversal invariant, order parameter.  If we include it in the low energy theory, 
\begin{align}
    \bar H=v\bk\cdot\boldsymbol{\sigma}\tau_z+\bB_v\cdot\boldsymbol{\sigma}\rho_z\tau_z+\bB\cdot\boldsymbol{\sigma}+
    (\Delta_0+\Delta\rho_x+\Delta_t\sigma_z\rho_z)\tau_x.
\end{align}

If the proximate superconductor has for example, it has two bands (pockets) related by time-reversal symmetry, it is possible to satisfy the relation $\Delta>\Delta_0$. Let us consider that superconductivity pairs two pockets by forming singlet states. Then we have,
\begin{align}
    \Delta_{sc}\propto\mu_x\sigma_0, \quad \lambda\propto 1, \quad\Rightarrow\quad\Delta_{ind}=\Delta\rho_x.
\end{align}
Here $\mu_z$ is a pocket index, and we considered the simplest coupling between the superconductor to the HOTI surface $\lambda$ (a $4\times 4$ identity matrix).  $\Delta\rho_x\tau_x$ is in this case the only induced order parameter, and $\Delta>\Delta_0$ is satisfied.  

Let us finally comment on the application of the previous logic to the surface of a HOTI, rather than a stand-alone two-dimensional system.  The effective two-dimensional surface Hamiltonian can be found by a projection of the three-dimensional bulk Hamiltonian, by a projector $P$, which we do not need to specify. The surface Hamiltonian is given by $PHP$, and the surface states are both eigenstates of $H$ and $P$. Applying the same procedure to the induced pairing terms, we can generally write 
\begin{align}
    \bar H_{eff}=PHP\tau_z-P\Delta_{ind} P .
\end{align}
Using the fact that a projector $P$ must satisfy $P=P^T$, it follows that $P\Delta_{ind} P\propto P^T\lambda^T\Delta_{sc}\inv\lambda P\tau_x\propto(\lambda P)^T(\lambda P)\tau_x$, which is also positive semi-definite, and thus it is subjected to no-go theorem of Ref.\cite{Haim16}.

\section{D. Application to known materials}
The effective field at each $\bB_v$ is a result of surface details, particularly how the bulk crystalline symmetry is broken at the surface. Let us comment on two important cases: When a HOTI is protected by inversion such as Bismuth \cite{Schindler18b}, $\bB_v$ has a domain wall in all its components. The surface magnitude $\bB_v$ is related to the Rashba energy, which can be very large for heavy materials. When close to the hinge, not only $m(y)$ but also $vk_0(y)$ becomes small, eventually changing sign. $\bB_v$ can reach magnitudes comparable to both the induced superconducting and Zeeman energies ($\sim\!1\rm meV$ with an applied field $\sim\!1\rm T$ for Bismuth\cite{Murani2017}). Ultimately, splitting or gapping of the hinge modes depends on a balance between these parameters and their spatial inhomogeneity. Note that the hinge states in Bismuth also coexist with other surface states and the low-energy Hamiltonian is not optimal. 

Now we comment on SnTe, another HOTI candidate. Here the surface gap can be controlled by the amount of strain that breaks the mirror symmetry \cite{Schindler18a}. The gap varies reaching as much as $60\rm meV$ for $5\%$ strain along the (110) direction. In SnTe, on the other hand $vk_0$ is very large and fairly independent of strain. While $vk_0$ poses a strict barrier to a split into one-dimensional Majorana modes and a topological superconductor is unlikely, SnTe is a very good candidate to realize a surface-based anomalous quantum Hall insulator. With a very large $g$-factor $\sim 50$ \cite{BERNICK1970569}, SnTe can invert the gap of one surface valley by combining a small amount of strain $<1\%$ creating a gap $\sim 10\rm meV$ with a applied magnetic field of reasonable magnitude $\sim10\rm T$.

\end{document}